\documentclass[]{aa}
\usepackage{txfonts}
\usepackage{graphicx}
\usepackage{natbib}
\bibpunct{(}{)}{;}{a}{}{,} 

\begin{document}

\title{\textsc{SAtlas}: Spherical Versions of the \textsc{Atlas} 
Stellar Atmosphere Program}

\author{John B. Lester \inst{1} \and 
Hilding R. Neilson \inst{2} }

\institute{Department of Chemical and Physical Sciences, 
University of Toronto Mississauga, lester@astro.utoronto.ca \and
Department of Astronomy and Astrophysics, University of Toronto, 
neilson@astro.utoronto.ca}

\date{Received 16 July 2008/ Accepted 30 August 2008}

\abstract
{The current stellar atmosphere programs still cannot 
match some fundamental observations of the brightest stars, and with 
new techniques, such as optical interferometry, providing new data for 
these stars, additional development of stellar atmosphere codes is 
required.}
{To modify the open-source model atmosphere program \textsc{Atlas} to 
treat spherical geometry, creating a test-bed stellar atmosphere 
code for stars with extended atmospheres.}
{The plane-parallel \textsc{Atlas} has been 
changed by introducing the necessary spherical modifications in the 
pressure structure, in the radiative transfer and in the temperature 
correction.}
{Several test models show that the spherical program 
matches the plane-parallel models in the high surface gravity regime, 
and matches spherical models computed by \textsc{Phoenix} and by 
\textsc{MARCS} in the low gravity case.}
{}

\keywords{Stars:atmospheres}

\authorrunning{J. B. Lester \& H. R. Neilson}
\titlerunning{Spherical Atlas}

\maketitle

\section{Introduction}

\textsc{Atlas} \citep{1970SAOSR.309.....K, 1979ApJS...40....1K} is a 
well-documented, well-tested, robust, open-source computer program for 
computing static, plane-parallel stellar atmospheres in local 
thermodynamic equilibrium (LTE).  Since \textsc{Atlas} came to maturity 
in the 1970s, stellar atmosphere codes have progressed in a number of 
directions to include important additional physics.  For example, the 
\textsc{Phoenix} program \citep{1999ApJ...512..377H} includes advances 
such as the massive use of statistical equilibrium (NLTE) in place of 
LTE, spherically symmetric extension in place of plane-parallel 
geometry, the inclusion of the dust opacities needed for brown dwarf 
temperatures, and the ability to include blast wave velocity fields.

These advances, while obviously moving toward greater reality, have not 
eliminated some persistent problems.  For example, a detailed study of 
Arcturus using \textsc{Phoenix} models \citep{2003ApJ...596..501S} 
found that their spherical NLTE models \textit{increased} the 
discrepancy between the observed and computed spectral irradiance.  A 
similar analysis of models with solar parameters 
\citep{2005ApJ...618..926S} also concluded that important opacity 
and/or other physics is still missing.  

This evidence that the significant improvements contained in the 
state-of-the-art programs have not achieved closer agreement with the 
observations of these fundamental, bright stars indicates a need to 
continue exploring additional physics.  This is a valuable role that 
\textsc{Atlas} can play because it is open source and freely available 
from Kurucz's web page (\textit{http://kurucz.harvard.edu}), where 
anyone can download the code and use it as the starting point for 
studying other possible improvements.  An example of the advantage 
provided by this starting point is the work of 
\citet{1996ASPC..108...73K} who explored convection as represented by 
the full spectrum of turbulence 
\citep{1991ApJ...370..295C, 1996ApJ...473..550C} as an alternative to 
the standard mixing length theory.  More recently, 
\citet{2007IAUS..239...71S} have developed a GNU-Linux port of 
\textsc{Atlas} for use in a range of applications.  In that spirit, we 
have developed versions of the \textsc{Atlas} program that replace the 
assumption of plane-parallel geometry by spherical symmetry.  These 
codes, comparable to the LTE, spherical version of \textsc{Phoenix} or 
of the spherical version of \textsc{MARCS} \citep{2008A&A...486..951G}, 
can serve as the basis for the study of additional physics needed to 
understand luminous stars that are cool enough that NLTE effects are 
not dominant.  Such continued studies are certainly necessary given the 
revolutionary achievements of optical interferometry in imaging the 
surfaces of stars, thus providing powerful new observational tests of 
models of stellar atmospheres.

\section{From \textsc{Atlas9} and \textsc{Atlas12} to 
\textsc{Atlas\_ODF} and \textsc{Atlas\_OS}}

In addition to including a wide range of continuous opacities, 
\textsc{Atlas} also incorporates the opacity of tens of millions of 
ionic, atomic and molecular lines.  The original treatment of these 
lines was via pre-computed opacity distribution functions (ODF) in the 
\textsc{Atlas9} program.  Later, \textsc{Atlas12} was developed to use 
opacity sampling (OS) of the same extensive line lists, and Kurucz 
continues to expand and improve the opacities for these codes. There 
are small parts of the two \textsc{Atlas} codes that must be different 
to handle the different line treatments, but the majority of the 
codes are not affected by the line treatment, and these can be 
identical.  However, as inevitably happens, small differences between 
the two codes develop over time; these can be seen by differencing the 
two source files.  Therefore, before beginning our conversion to 
spherical geometry, we rationalized the two versions of \textsc{Atlas} 
to be identical in every way except where they must be different for 
the line treatments. Where there are differences not associated with 
the treatment of the line opacity, we have adopted the \textsc{Atlas12}
routine, under the assumption that it is likely to be newer and 
undergoing more active development than the older \textsc{Atlas9} code.
We have also converted our codes to current Fortran 2003 standards.  To 
distinguish our programs from the Kurucz originals, we use the name 
\textsc{Atlas\_ODF} for our version of \textsc{Atlas9}, and 
\textsc{Atlas\_OS} for our version of \textsc{Atlas12}.

\begin{figure}
 \resizebox{\hsize}{!}{\includegraphics{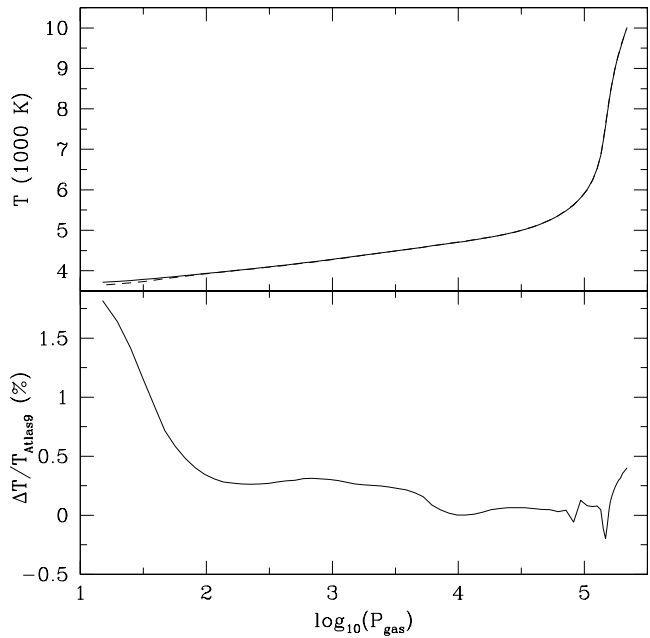}}
 \caption{\label{fig:tp_odf_kur} 
 A comparison of the temperature structures for two solar atmospheric 
 models.  The top panel displays the temperatures as a function of 
 $\log_{10} (P_{\mathrm{gas}})$.  The solid line is the model computed 
 with \textsc{Atlas\_ODF} and the dashed line is the solar model 
 atmosphere \texttt{asunodfnew.dat} from the Kurucz web page.  
 The bottom panel shows the percentage difference between the 
 temperatures of the two models, also as a function of the gas pressure.
 The differences are entirely due to the conversion to Fortran 2003 and 
 to using consistent, modern values of fundamental parameters 
 throughout the \textsc{Atlas\_ODF} code.  The numerical methods are 
 the same for the two models.
 }
\end{figure}

To test the accuracy of our conversions, we computed a model of the 
solar atmosphere, starting from the input model \texttt{asunodfnew.dat} 
dated 2 April 2008 in the Kurucz directory 
\texttt{/grids/gridp00odfnew}.  We learned later that this model was 
computed by Castelli as part of her collaboration with Kurucz
\citep{2004astro.ph..5087C}.  We used the opacity distribution function 
\texttt{bdfp00fbig2.dat} located Kurucz's directory 
\texttt{/opacities/dfp00f}, which uses the solar iron abundance 
$= -4.51$.  After 20 iterations, the flux errors are all $\leq 0.2\%$ 
and the flux derivative errors are all $\leq 2.8\%$; most errors of 
both kinds are smaller than our output resolution of 0.1\%.  
Figure~\ref{fig:tp_odf_kur} compares the temperature and pressure 
structures of the solar model computed with \textsc{Atlas\_ODF} and the 
starting \textsc{Atlas9} model.

Starting from the \textsc{Atlas\_ODF} code, we removed those 
components that used the opacity distribution functions and replaced 
them with the components needed to do opacity sampling.  At this point 
we introduced several changes that are not present in \textsc{Atlas12}.
Depending on the effective temperature of the star, \textsc{Atlas12} 
adjusts the index of the starting wavelength, variable \texttt{nustart},
to eliminate wavelengths were the flux is negligibly small.  In 
anticipation of future applications of the code, we have removed this 
adjustment to always begin at the shortest wavelength, independent of 
the effective temperature.  This change has been propagated back to the 
\textsc{Atlas\_ODF} version.  Second, \textsc{Atlas12} always computes 
30,000 wavelengths with a wavelength spacing equal to a constant 
spectral resolving power of $10^4$.  We have modified this to be able 
to specify a spectral resolving power $\leq 10^4$, and have the number 
of wavelengths adjust automatically.  We did this to test the 
dependence of the computed model on the spectral resolving power.  
Third, in assembling the master file of lines to be sampled, 
\textsc{Atlas12} uses a sorting routine from the computer's operating 
system, which is outside of the source code.  We have modified the 
subroutine \texttt{prelinop}, to perform this sort within the 
\textsc{Atlas} source code, making it self-contained.

\begin{figure}
 \resizebox{\hsize}{!}{\includegraphics{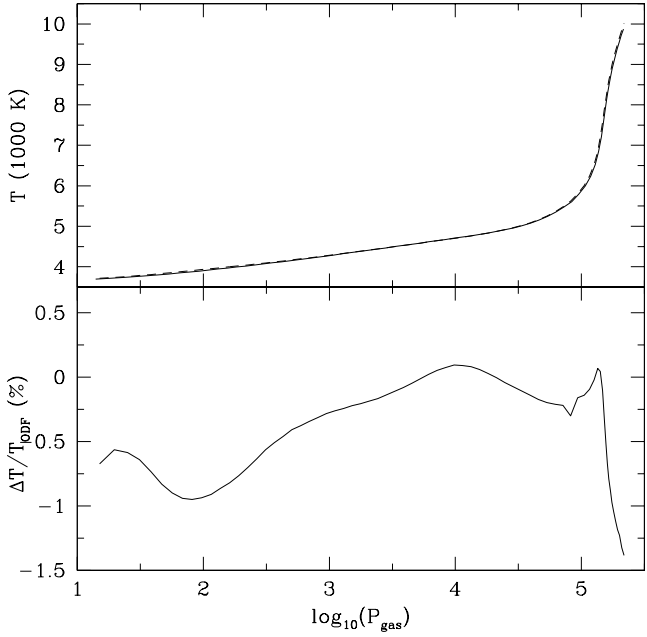}}
 \caption{\label{fig:tp_os_odf}
 A comparison of the temperature structures for two solar atmospheric 
 models. The top panel displays the temperatures as a function of 
 $\log_{10} (P_{\mathrm{gas}})$.  The solid line is for the model 
 computed with \textsc{Atlas\_OS} and the dashed line is the model 
 computed with \textsc{Atlas\_ODF}.  
 The bottom panel shows the percentage difference between the 
 temperatures of the two models, also as a function of the gas pressure.
 The differences are entirely due to the way in which line opacity is 
 included.  Both codes use Fortran 2003, the same fundamental 
 parameters and the same numerical methods.
 }
\end{figure}

To test \textsc{Atlas\_OS}, we computed a model of the solar atmosphere,
again starting from the input model \texttt{asunodfnew.dat}.  We used 
Kurucz's files \texttt{lowlines}, \texttt{hilines} and 
\texttt{bnltelines8} for the atomic and ionic lines, as well as the 
molecular files \texttt{diatomic}, \texttt{tiolines} and 
\texttt{h2ofast}.  After eliminating lines that did not contribute at 
the temperatures of the solar atmosphere, the number of sampled lines 
used in the calculation was about two million.  
Figure~\ref{fig:tp_os_odf} compares the temperature structure of the 
solar model computed with \textsc{Atlas\_OS} and the model computed 
with \textsc{Atlas\_ODF}.  The differences between the models using 
the two methods of including line opacity are comparable to the 
differences between our \textsc{Atlas\_ODF} and the original 
\textsc{Atlas9} shown in Fig.~\ref{fig:tp_odf_kur}.  Therefore, the 
joint agreements displayed in Fig.~\ref{fig:tp_odf_kur} and 
Fig.~\ref{fig:tp_os_odf} show that our \textsc{Atlas\_OS} code matches 
the \textsc{Atlas9} structure.  As an additional test of the two line 
treatments, we have used our modification to the opacity-sampled code 
mentioned earlier to vary the spectral resolving power.  The 
opacity-sampled model shown in Fig.~\ref{fig:tp_os_odf} used the 
default spectral resolving power of 30,000.  Our tests found that 
repeating the calculation with spectral resolving powers of 10,000 and 
3,000 produced almost no change in the resulting model structure.

\begin{figure}
 \resizebox{\hsize}{!}{\includegraphics{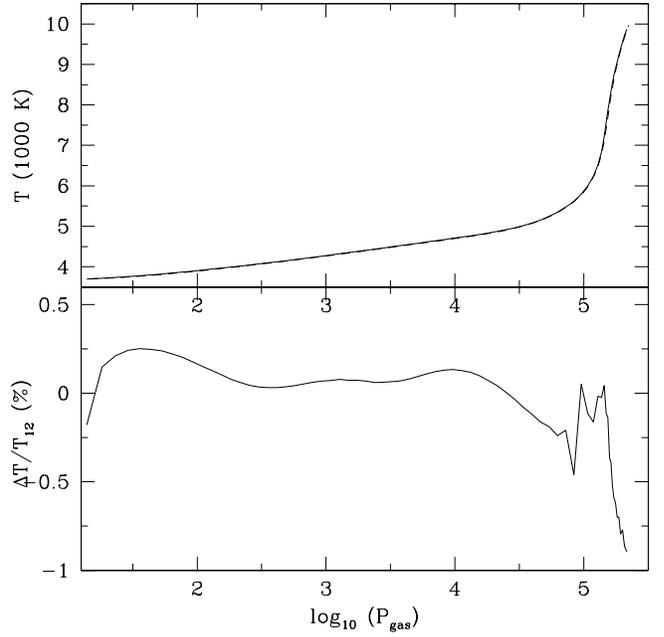}}
 \caption{\label{fig:tp_os_kur08}
 A comparison of the temperature structures for two solar atmospheric 
 models.  The top panel displays the temperature as a function of 
 $\log_{10} (P_{\mathrm{gas}})$. The solid line is the model computed 
 with \textsc{Atlas\_OS}, and the dashed line is an \textsc{Atlas12} 
 model provided by Kurucz (private communication).
 The bottom panel shows the percentage difference between the 
 temperatures of the two models, also as a function of the gas pressure.
 }
\end{figure}

As an additional test, Kurucz (private communication) provided us with 
a new \textsc{Atlas12} (opacity-sampled) solar model that we compare 
with our \textsc{Atlas\_OS} model in Fig.~\ref{fig:tp_os_kur08}. It is 
apparent that the agreement is very good.

\section{From \textsc{Atlas\_ODF} to \textsc{SAtlas\_ODF}}

The spherically symmetric version of the code, \textsc{SAtlas\_ODF}, 
was created from \textsc{Atlas\_ODF}.  Plane-parallel models are 
labeled with the parameters effective temperature, $T_{\mathrm{eff}}$, 
and surface gravity, usually given as $\log g$ in cgs units.  For 
spherical models these two parameters are degenerate because the same 
value of $\log g$ is produced by different combinations of the stellar 
mass and radius.  Therefore, we have elected to use the three 
fundamental physical parameters luminosity, $L_{\ast}$, mass, 
$M_{\ast}$, and radius, $R_{\ast}$.  These can be supplied in cgs units 
or as multiples or fractions of the solar values, $L_{\ast}/L_{\sun}$, 
$M_{\ast}/M_{\sun}$ and $R_{\ast}/R_{\sun}$, where the values of 
$L_{\sun}$, $M_{\sun}$ and $R_{\sun}$ are available throughout the 
source code via a Fortran 2003 module routine.

The radius, of course, will vary with depth in the extended atmosphere.
Therefore we have chosen to define the radius where the radial 
Rosseland mean optical depth, $\tau_{\mathrm{R}}$, has the value of 
$2/3$ because a photon has a probability of about 50\% of escaping the 
atmosphere from that depth.  Other choices could be made, such as 
$\tau_{\mathrm{R}} = 1$ \citep{2008A&A...486..951G}, or $\tau_{500} = 1$
\citep{1999ApJ...512..377H}, but these differences are nearly 
negligible.

With the three basic parameters defined, there are three modifications 
to the code: pressure structure, radiative transfer and temperature 
correction.

\subsection{Pressure structure}

\textsc{Atlas9} and \textsc{Atlas12} both solve for the pressure 
structure in two locations.  After reading in the starting model, the 
pressure structure is solved from the initial temperature structure 
as a function of Rosseland mean optical depth, $T(\tau_{\mathrm{R}})$, 
in the subroutine \texttt{ttaup}.  After the first iteration the total 
gas pressure is computed by integrating the simple equation 
\begin{equation}
\textrm{d} P_{\mathrm{tot}}/\textrm{d}m = g,
\end{equation}
where $m$ is the mass column density defined as
\begin{equation}
\label{eq:dm}
\textrm{d}m = - \rho \textrm{d}r,
\end{equation}
and $g$ is the constant gravitational acceleration in the 
plane-parallel atmosphere.

In a spherical atmosphere there are three potential depth variables: 
mass column density, $m$, radius, $r$, and \emph{radial} Rosseland 
mean optical depth, $\tau_{\mathrm{R}}$.  As discussed by 
\citet{1974ApJS...28..343M}, there is no clear preference for any of 
these variables.  Therefore, we have elected to adopt the radial 
Rosseland mean optical depth as our primary variable, and then use 
$T(\tau_{\mathrm{R}})$ to solve for the pressure structure by 
modifying the subroutine \texttt{ttaup}.  This is done in the 
initialization of the calculation and for each iteration.

The modifications to the subroutine \texttt{ttaup} include allowing the 
surface gravity to vary with the radial distance from the center of 
the star, 
\begin{equation}
g(r) = G \frac{M(r)}{r^2}.
\end{equation}
Because the mass of the atmosphere is negligible compared to the mass 
of the star, it is an excellent approximation to set 
$M(r) = M_{\ast}$, giving
\begin{equation}
g(r) = G \frac{M_{\ast}}{r^2}.
\end{equation}
If the starting model is a spherical model, there will be initial 
values for $r$.  If the starting model is plane parallel, we solve 
for $r$ from the defining differential equation
\begin{equation}
\textrm{d} r = - \frac{1}{\rho k_{\mathrm{R}}(r)} \textrm{d} \tau_{\mathrm{R}},
\end{equation}
in its logarithmic form to minimize numerical errors.  Here $\rho(r)$ 
is the mass density, and $k_{\mathrm{R}}(r)$ is the Rosseland mean 
opacity, the sum of absorption, $\kappa$, and scattering, $\sigma$, per 
unit mass as a function of depth, both of which are available from the 
input model.  This solution begins by assuming an initial value for the 
atmosphere's extension, $r(1)/R_{\ast}$, after which the solution is 
performed using the Bulirsch-Stoer method given in 
\textit{Numerical Recipes in Fortran 90} \citep{1996nrfa.book.....P}.  
The fifth-order Runga-Kutta method, also from \textit{Numerical Recipes 
in Fortran 90}, was also investigated, and is built into the source 
code, but we found the results from the two methods to be identical.  
After finishing the solution for $r$, the atmospheric extension, 
$r(1)/R_{\ast}$, is checked against the initial assumption.  If the 
extension differs by more than $10^{-6}$ from the starting assumption,
the starting assumption is updated and the solution is iterated until 
the extension converges to $< 10^{-6}$.

Using $g(r)$, the equation of spherical hydrostatic equilibrium is 
\begin{equation}
\label{eq:sph_hydro}
\frac{\textrm{d} P_{\mathrm{tot}}}{\textrm{d} \tau_{\mathrm{R}}} 
= \frac{g(r)}{k_{\mathrm{R}}(r)},
\end{equation}
The solution begins at the upper boundary by assuming an initial value 
of $k_{\mathrm{R}}(1)$.  Then, following 
\citet{1974ApJS...28..343M}, we assume all properties, except pressure 
and density, are constant above $r(1)$, and that these variables 
decrease with a constant scale height.  This leads to
\begin{equation}
P_{\mathrm{tot}}(1) 
= g(1) \frac{\tau_{\mathrm{R}}(1)}{k_{\mathrm{R}}(1)}.
\end{equation}
The gas pressure, $P_{\mathrm{g}}(1)$, is derived from the total 
pressure by subtracting values for radiation pressure, 
$P_{\mathrm{r}}(1)$, and turbulent pressure, $P_{\mathrm{t}}(1)$, if 
these are known.  The gas pressure and the temperature are then used to 
interpolate an updated value for $k_{\mathrm{R}}(1)$ from the input 
model.  This procedure is iterated until the upper boundary pressure 
converges to $< 10^{-6}$.

With the upper boundary condition established, Eq.~\ref{eq:sph_hydro}
is integrated for $P_{\mathrm{tot}}$, again using the Bulirsch-Stoer 
method.  At each step the gas pressure is found as described above, 
and the gas pressure and temperature are used to interpolate the 
corresponding value for the Rosseland mean opacity.

This method of solving the hydrostatic equilibrium is also applicable 
to the plane-parallel atmosphere with $g(r) = g$ and without solving for
the radius.  To test our implementation, we have incorporated the 
modified version of subroutine \texttt{ttaup}, with both the 
Bulirsh-Stoer and the fifth-order Runga-Kutta routines, into 
\textsc{Atlas\_ODF} and \textsc{Atlas\_OS}.  The maximum difference 
between the Bulirsch-Stoer (or the fifth-order Runga-Kutta) method and 
the original Hamming method was less than our output numerical 
resolution of 1 part in $10^4$ at all but two of the 72 depth points. 
Therefore, the percentage difference between the methods is zero except 
at these two depths, where the differences are only $0.021\%$ and 
$0.014\%$.  It is clear that the pressure solution is being done 
correctly.

\subsection{Radiative Transfer}

\textsc{Atlas9} and \textsc{Atlas12} solve the radiative transfer using 
the integral equation method.  The complication introduced by a 
geometrically extended, spherically symmetric atmosphere is that the 
angle between a ray of light and the radial direction varies with 
depth.  Numerous methods are available for solving this problem, of 
which we have chosen to use the \citet{1971JQSRT..11..589R} 
reorganization of the \citet{1964CR...258..3189F} method.

\begin{figure}
 \resizebox{\hsize}{!}{\includegraphics{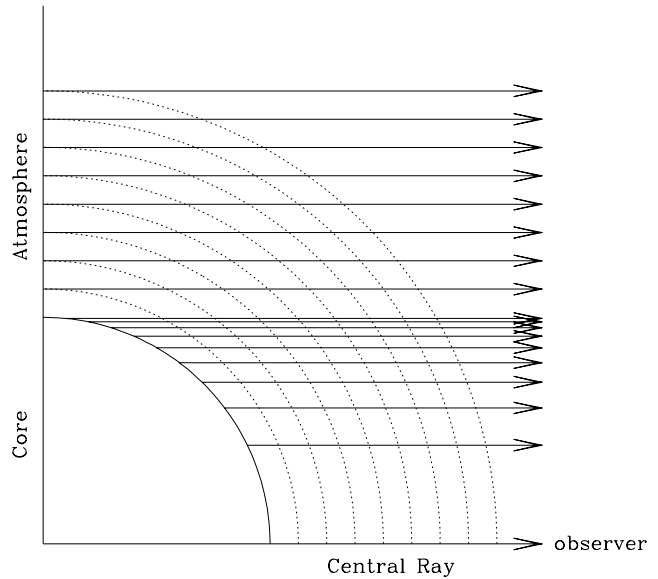}}
 \caption{\label{fig:sph_geo}
 The geometry used for the \citet{1971JQSRT..11..589R} method of 
 radiative transfer.  The rays are distributed over the core in equal 
 steps of 0.1 in $\mu$.}
\end{figure}

Following the approach described by \citet{1978stat.book.....M}, we 
solve the radiative transfer along a set of rays parallel to the 
central ray directed toward the distant observer, as shown in 
Fig.~\ref{fig:sph_geo}.  A subset of these rays intersect the 
``core'' of the star, defined as the deepest radial optical depth, 
which we usually set to be $\tau_{\mathrm{R}} = 100$.  We sample the 
surface of the core using 10 rays.  We tried both equal steps of 
$\mu = \cos \theta$ covering the interval $1.0 \leq \mu \leq 0.1$ in 
steps of $\Delta \mu = 0.1$, which is shown in Fig.~\ref{fig:sph_geo},
and a finer spacing toward the edge of the core by distributing the 
rays as $\mu = 1.0, 0.85, 0.7, 0.55, 0.4, 0.25, 0.2, 0.15, 0.1, 0.05$.
We found the results to be almost identical, so we chose to use the 
equal $\mu$ spacing.
For the core rays, the lower boundary condition of the radiative 
transfer is the diffusion approximation.  From the core to the 
surface we follow the convention used by Kurucz in his plane-parallel 
models by having 72 depth points.  For the central ray these depth 
points are distributed eight per decade with equal steps of the 
$\Delta \log \tau_{\mathrm{R}} = 0.125$ from 
$\log \tau_{\mathrm{R}} = 2$ to -6.875.  For the off-center core rays 
the steps will be different, depending on the projection.

The tangent rays are those that terminate at the radius perpendicular 
to the central ray, and which are tangent to a particular atmospheric 
shell at that point, as shown in Fig.~\ref{fig:sph_geo}.  The 
\emph{radial} spacing between the shells is set by the central ray, and 
these spacings define the impact parameters of all the tangent rays.   
With this geometry, we calculated values of $\mu$ at the intersection 
of each ray toward the distant observer with each atmospheric depth, 
and from these we compute the integration weights at each point over 
the surface of each atmospheric shell at every depth.  The lower 
boundary condition for the radiative transfer of the tangent rays is 
the assumption of symmetry at the perpendicular radius. At the surface 
of the atmosphere the rays toward the distant observer have $\mu$ 
values that depend on the steps described above.  When we want to use 
these surface intensities, such as to predict the observable 
center-to-limb variation, we map the computed $I(\mu)$ onto a specified 
set of $\mu$-values using a cubic spline interpolation.

\begin{figure}
 \resizebox{\hsize}{!}{\includegraphics{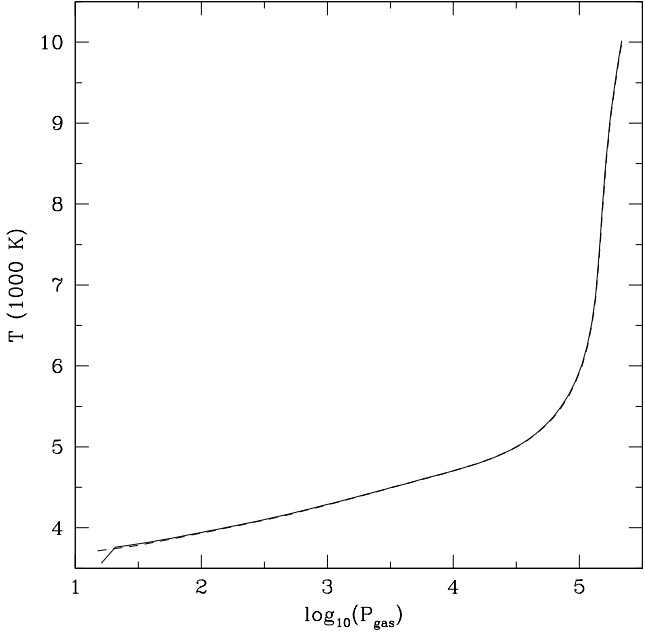}}
 \caption{\label{fig:tp_ryb_josh}
 A comparison of the temperature structures for two solar atmospheric 
 models, both using the \textsc{Atlas\_ODF} code.  The solid line used 
 the \citet{1971JQSRT..11..589R} method to compute the radiative 
 transfer, and the dashed line used the original \textsc{Atlas9} 
 integral equation method.}
\end{figure}

This solution with the \citet{1971JQSRT..11..589R} organization uses 
\emph{exactly} the same equations as the original 
\citet{1964CR...258..3189F} method.  Therefore, in the plane-parallel 
case, in which both can be used, the results must be exactly the same.
To test this, we created two alternative radiative transfer routines 
for the plane-parallel codes \textsc{Atlas\_ODF} and 
\textsc{Atlas\_OS}, one new routine having the original Feautrier 
organization and the other having the Rybicki organization, and 
we select which of the radiative transfer routines we want by using an 
input instruction at run time, holding everything else the same.  The 
result of the comparison is shown in Fig.~\ref{fig:tp_ryb_josh}.
Using the \citet{1964CR...258..3189F} organization gives exactly the 
same result (the output files have zero differences), as it should. 
The tiny temperature drop at the top of the atmosphere shown in 
Fig.~\ref{fig:tp_ryb_josh} is entirely due to the precise form of 
implementing the surface boundary condition.  We explored different 
implementations (changing one statement) that were logically 
equivalent, and we found the result shown to be the closest match to 
the original \textsc{Atlas} integral equation solution. The other 
implementations gave the same temperature at the top depth of the 
atmosphere, $T(\tau_1)$, but have a slower convergence to the 
temperature derived using the integral equation method.

A note about the relative run times of the same code using the three 
different radiative transfer routines: the time per iteration using the
Feautrier method is about half the time of the original integral 
equation method, while with the Rybicki method the time per iteration 
is about ten times longer than the original integral equation method.

The difference in the execution time of the Feautrier and the Rybicki 
methods, which use the same equations, is due to the sizes of the 
matrices that must be inverted.  The Feautrier method computes the 
radiation field for all $\mu$ values at each atmospheric depth, where 
the $\mu$ values are the double-Gauss angles found to be superior by 
\citet{1951MNRAS...111..377S}.  The computing time to invert the 
matrices scales as the cube of the number of $\mu$ values, $M^3$.  We 
use three angle points because our tests using four and eight angle 
points are insignificantly different ($\Delta T < 1$ K) from the 
three-angle solution, while the computing time is certainly increased.

The Rybicki method computes the radiation for each individual ray over 
all atmospheric depths. The number of depth points ranges from just two 
for the tangent ray that penetrates just one atmospheric depth, up to 
72 depth points for the rays that reach the core.  The need to invert 
these larger matrices causes the execution of the Rybicki method to be 
longer.

\subsection{Temperature Correction}

\textsc{Atlas9} and \textsc{Atlas12} perform the temperature correction 
using the Avrett-Krook method \citep{1963ApJ...137..874A, 
1964SAOSR.167...83A} modified to include convection 
\citep{1970SAOSR.309.....K}.  While the other spherical atmosphere 
programs (\textsc{Phoenix} and \textsc{MARCS}) use their own methods 
to perform temperature corrections, it is our experience 
\citep{2002PASP..114..330T} that the Avrett-Krook method is extremely 
robust, capable of achieving temperature convergence $\leq 1$ mK.  
Therefore, we have chosen to modify the Avrett-Krook temperature 
correction routine in the original \textsc{Atlas} codes to include 
spherically symmetric extension.   

We start with the time-independent equation of radiative transfer in 
spherical geometry \citep{1978stat.book.....M},
\begin{equation}
\label{eq:sph_rad_tran}
\mu \frac{\partial I(\nu)}{\partial r} 
+ \frac{1 - \mu^2}{r} \frac{\partial I(\nu)}{\partial \mu}
= k(\nu) [S(\nu) - I(\nu)],
\end{equation}
where $S(\nu)$ is the source function given by 
\begin{equation}
\label{eq:source_function}
S(\nu, T) = \sigma(\nu) J(\nu) + [1 - \sigma(\nu)] B(\nu, T).
\end{equation}
To be consistent with the approach in \textsc{Atlas}, we express 
Eq.~\ref{eq:sph_rad_tran} in terms of the mass column density 
(Eq.~\ref{eq:dm}) to obtain
\begin{equation}
\label{eq:sph_rad_tran_m}
- \rho \mu \frac{\partial I(\nu)}{\partial m} 
+ \frac{1 - \mu^2}{r} \frac{\partial I(\nu)}{\partial \mu}
= k(\nu) [S(\nu) - I(\nu)].
\end{equation}

In general, Eq.~\ref{eq:sph_rad_tran_m} does \emph{not} conserve the 
luminosity with depth because the atmospheric temperature structure is 
wrong, but we assume that small perturbations to the current structure 
will produce a constant luminosity with depth.  That is, we introduce 
the perturbations
\begin{equation}
\label{eq:perturb_m}
m = m_0 + \delta m,
\end{equation}
\begin{equation}
\label{eq:perturb_r}
r = r_0 + \delta r,
\end{equation}
\begin{equation}
\label{eq:perturb_t}
T = T_0 + \delta T,
\end{equation}
and
\begin{equation}
\label{eq:perturb_i}
I(\nu) = I_0(\nu) + \delta I(\nu),
\end{equation}
where the subscript 0 refers to the current structure.  We also use the 
subscript 0 for the current extinction, $k_0(\nu)$, and the current 
source function, $S_0(\nu)$.

The extinction and the source function are expanded to first order as 
\begin{equation}
\label{eq:expand_k}
k(\nu) = k_0(\nu) + \frac{\partial k_0(\nu)}{\partial m} \cdot \delta m,
\end{equation}
and 
\begin{equation}
\label{eq:expand_s}
S(\nu) = S_0(\nu) + \frac{\partial S_0(\nu)}{\partial m} \cdot \delta m 
           + \frac{\partial S_0(\nu)}{\partial T} \cdot \delta T.
\end{equation}
To simplify the notation further, we represent 
$\partial k_0(\nu) / \partial m = k_0^\prime(\nu)$,
$\partial S_0(\nu) / \partial m = S_0^\prime(\nu)$,
and 
$\partial S_0(\nu) / \partial T = \dot{S}_0(\nu)$.

Using Eq.~\ref{eq:perturb_m} we can write 
\begin{equation}
\label{eq:dm_1}
\frac{\textrm{d} m}{\textrm{d} m_0} =
 1 + \frac{\textrm{d} \delta m}{\textrm{d} m_0}
\end{equation}
or
\begin{equation}
\label{eq:dm_1_2}
\textrm{d} m = \textrm{d} m_0(1 + \delta m^{\prime}),
\end{equation}
where
\begin{equation}
\label{eq:dm_prime}
\delta m^{\prime} = \frac{\textrm{d} \delta m}{\textrm{d} m_0}.
\end{equation}
Therefore, the derivative in the first term in 
Eq.~\ref{eq:sph_rad_tran_m} becomes
\begin{equation}
\frac{\partial I(\nu)}{\partial m}
= 
\frac{\partial I(\nu)}{\partial m_0} (1 + \delta m^{\prime})^{-1}.
\end{equation}
In the second term in Eq.~\ref{eq:sph_rad_tran_m} there is 
$r^{-1} = (r_0 + \delta r)^{-1}$.  Because we assume that 
$\delta r \ll r_0$, we can perform a binomial expansion of $r^{-1}$, 
keeping only the first two terms, to get 
\begin{equation}
\label{eq:binom_r}
\frac{1}{r} = \frac{1}{r_0 + \delta r} \approx 
\frac{1}{r_0} \left (1 - \frac{\delta r}{r_0} \right )
= \frac{1}{r_0} \left (1 + \frac{\delta m}{\rho r_0} \right )
\end{equation}
by using Eq.~\ref{eq:dm}.

Using these, the spherical radiative transfer equation, including 
perturbations, becomes
\begin{eqnarray}
\label{eq:sph_rad_tran_pert}
\lefteqn{\frac{- \rho \mu}{1 + \delta m^\prime}\frac{\partial[I_0(\nu) +
\delta I(\nu)]} {\partial m_0} 
 + \frac{1 - \mu^2}{r_0} \left (1 + \frac{\delta m}{\rho r_0} \right )
  \frac{\partial [I_0(\nu) + \delta I(\nu)]}{\partial \mu} = }
 \nonumber \\
& & [k_0(\nu) + k_0^\prime(\nu) \cdot \delta m]
 [S_0(\nu) + S_0^\prime(\nu) \cdot \delta m 
           + \dot{S}_0(\nu) \cdot \delta T \nonumber \\
& & {}-  I_0(\nu) - \delta I(\nu)].
\end{eqnarray}
Clearing the $1/(1 + \delta m^{\prime})$ term and expanding 
Eq.~\ref{eq:sph_rad_tran_pert}, ignoring terms with second-order 
perturbations, gives
\begin{eqnarray}
\label{eq:sph_rad_tran_pert_expand}
\lefteqn{ - \rho \mu \frac{\partial I_0(\nu)}{\partial m_0}
 - \rho \mu \frac{\partial \delta I(\nu)}{\partial m_0}
 + \frac{1 - \mu^2}{r_0} \frac{\partial I_0(\nu)}{\partial \mu} 
 + \frac{1 - \mu^2}{r_0} \frac{\partial \delta I(\nu)}{\partial \mu} }
\nonumber \\
& & {} + \frac{1 - \mu^2}{r_0} \left ( \delta m^{\prime} +
                                 \frac{\delta m}{\rho r_0} \right )
                   \frac{\partial I_0(\nu)}{\partial \mu} = 
 k_0(\nu) [S_0(\nu) - I_0(\nu)] \nonumber \\
& & {} + [k_0(\nu) \cdot \delta m^\prime + k_0^\prime \cdot \delta m]
  [S_0(\nu) - I_0(\nu)] \nonumber \\
& & {} + k_0(\nu) [S_0^\prime(\nu) \cdot \delta m +
             \dot{S}_0(\nu) \cdot \delta T - \delta I(\nu)].
\end{eqnarray}
Note that the left hand side of Eq.~\ref{eq:sph_rad_tran_pert_expand} 
contains 
\begin{equation}
\label{eq:sph_rad_tran_0l}
- \rho \mu \frac{\partial I_0(\nu)}{\partial m_0}
+ \frac{1 - \mu^2}{r_0}\frac{\partial I_0(\nu)}{\partial \mu},
\end{equation}
and the right hand side has 
\begin{equation}
\label{eq:sph_rad_tran_0r}
k_0(\nu) [S_0(\nu) - I_0(\nu)],
\end{equation}
and these equal each other because they are just the two sides of 
Eq.~\ref{eq:sph_rad_tran_m} for the current structure.  Canceling these 
out of Eq.~\ref{eq:sph_rad_tran_pert_expand} leaves the first-order 
perturbation of the spherical equation of radiative transfer
\begin{eqnarray}
\label{eq:sph_rad_tran_1}
\lefteqn{- \rho \mu \frac{\partial \delta I(\nu)}{\partial m_0}
+ \frac{1 - \mu^2}{r_0}\frac{\partial \delta I(\nu)}{\partial \mu} 
+ \frac{1 - \mu^2}{r_0}
 \left (\delta m^{\prime} + \frac{\delta m}{\rho r_0} \right )
 \frac{\partial I_0(\nu)}{\partial \mu} =} \nonumber \\
& & [k_0(\nu) \cdot \delta m^{\prime} +  k_0^{\prime}(\nu) \cdot \delta m]
[S_0(\nu) - I_0(\nu)] \nonumber \\
& & {} + k_0(\nu) [S_0^{\prime}(\nu) \cdot \delta m +
            \dot{S_0}(\nu) \cdot \delta T - \delta I(\nu)].
\end{eqnarray}

The first angular moment of the first-order perturbation equation is 
obtained by multiplying Eq.~\ref{eq:sph_rad_tran_1} by $\mu$ and 
integrating over all $\mu$, to get
\begin{eqnarray}
\label{eq:sph_pert_moment_1}
\lefteqn{- \rho \frac{\partial \delta K(\nu)}{\partial m_0}
+ \frac{1}{r_0} [3 \delta K(\nu) - \delta J(\nu)] = } \nonumber \\
& & {} - [k_0(\nu) \cdot \delta m^{\prime} 
       +  k_0^\prime(\nu) \cdot \delta m] 
          H_0(\nu) - k_0(\nu) \delta H(\nu) \nonumber \\
& & {} - \frac{1}{r_0} \left (\delta m^{\prime} + 
                         \frac{\delta m}{\rho r_0} \right )
  [3K_0(\nu) - J_0(\nu)].
\end{eqnarray}
Dividing Eq.~\ref{eq:sph_pert_moment_1} by $k_0(\nu)$ and integrating 
over all frequencies, we obtain 
\begin{eqnarray}
\label{eq:sph_moment_1_int}
\lefteqn{\int_0^{\infty}\frac{1}{k_0(\nu)}
\left [ \frac{3 \delta K(\nu) - \delta J(\nu)}{r_0} -
        \rho \frac{\partial \delta K(\nu)}{\partial m_0} \right ] 
  \textrm{d} \nu =} \nonumber \\
& & {} - \delta m^{\prime} \int_0^{\infty} H_0(\nu) \textrm{d} \nu 
       - \delta m \int_0^{\infty} \frac{k_0^{\prime}(\nu)}{k_0(\nu) } 
          H_0(\nu) \textrm{d} \nu 
       - \int_0^{\infty} \delta H(\nu) \textrm{d} \nu \nonumber \\ 
& & {} - \frac{1}{r_0}
    \left (\delta m^{\prime} + \frac{\delta m}{\rho r_0} \right )
    \int_0^{\infty} \frac{[3K_0(\nu) - J_0(\nu)]}{k_0(\nu)} \textrm{d} \nu. 
\end{eqnarray}
We now assume that the correct choice of $\delta m$ will make the 
left hand side of Eq.~\ref{eq:sph_moment_1_int} go to zero.  That is, 
we assume that the perturbations of the radiation field, $\delta K$ and
$\delta J$, vanish when the correct atmospheric structure is obtained.  
These assumptions are equivalent to the assumptions used in 
\citet{1964SAOSR.167...83A}, equation 25, and in 
\citet{1970SAOSR.309.....K}, equation 7.5.  This leaves the right hand 
side of Eq.~\ref{eq:sph_moment_1_int} as a differential equation for 
$\delta m$,
\begin{equation}
\label{eq:delta_r_deq}
a_0 \delta m^{\prime} + b_0 \delta m + c_0 = 0,
\end{equation}
where
\begin{equation}
\label{eq:delta_r_deq_a}
a_0 = H_0 + 
\frac{1}{r_0} \int_0^{\infty} \frac{[3K_0(\nu) - J_0(\nu)]}
                                   {k_0(\nu)} \textrm{d} \nu,
\end{equation}
\begin{equation}
\label{eq:delta_r_deq_b}
b_0 = \int_0^{\infty} \frac{k_0^{\prime}(\nu)}{k_0(\nu)}H_0(\nu) \textrm{d} \nu
    + \frac{1}{\rho r_0^2} \int_0^{\infty} \frac{[3K_0(\nu) - J_0(\nu)]}
                                           {k_0(\nu)} \textrm{d} \nu
\end{equation}
and
\begin{equation}
\label{eq:delta_r_deq_c}
c_0 = \int_0^{\infty} \delta H(\nu) \textrm{d} \nu = 
\delta H = \mathcal{H} - H_0,
\end{equation}
with 
\begin{equation}
\mathcal{H} = \frac{L_{\ast}}{(4 \pi r)^2}
\end{equation}
being the radially dependent Eddington flux that we need to achieve.  
The general solution to Eq.~\ref{eq:delta_r_deq} is 
\begin{equation}
\label{eq:delta_r_deq_sol}
\delta m = - \exp \left [-\int \frac{b_0(\tilde{m})}
                                    {a_0(\tilde{m})} 
                           \textrm{d} \tilde{m} \right ]
\int \frac{c_0(\tilde{m})}{a_0(\tilde{m})} 
e^{\int \frac{b_0(\tilde{m})}{a_0(\tilde{m})} \mathrm{d} \tilde{m}} 
\textrm{d} \tilde{m},
\end{equation}
where $\tilde{m}$ is an integration variable.

The correction for the mass column density found above has assumed that 
all the energy is carried by radiation.  If the atmospheric temperature 
is cool enough, significant amounts of energy can also be carried by 
convection in the deeper, less transparent levels of the atmosphere.  
\textsc{Atlas} calculates the convective energy transport by the mixing 
length approximation.  The equations in \citet{1970SAOSR.309.....K} 
do not contain the radial variable explicitly, but they do contain 
the surface gravity, $g$, which now varies with $r$.  However, the 
implementation of those equations replaces $g$ in terms of the total 
pressure, which now \emph{implicitly} includes the geometry.  Therefore,
there is no need to modify the original \textsc{Atlas} code to include 
convection in the spherical temperature correction, and 
Eq.~\ref{eq:delta_r_deq} remains the same, with the addition of 
convective terms in the coefficients $a_0, \ b_0$ and $c_0$ as follows:
\begin{eqnarray}
\label{eq:delta_r_deq_ac}
a_0 & = & H_0(\mathrm{rad})
+ \frac{1}{r_0} \int_0^{\infty} \frac{[3K_0(\nu) - J_0(\nu)]}
                {k_0(\nu)} \textrm{d} \nu \nonumber \\
& & {} + H_0(\mathrm{conv}) \frac{3 \nabla}{2(\nabla - \nabla_{\mathrm{ad}})}
  \left (1 + \frac{D}{D + \nabla - \nabla_{\mathrm{ad}}} \right )
\end{eqnarray}
\begin{eqnarray}
\label{eq:delta_r_deq_bc}
\lefteqn{b_0 = \int_0^{\infty} \frac{k_0^{\prime}(\nu)}{k_0(\nu)}H_0(\nu) \textrm{d} \nu 
+ \frac{1}{\rho r_0^2} \int_0^{\infty} \frac{[3K_0(\nu) - J_0(\nu)]}
                                            {k_0(\nu)} \textrm{d} \nu } \nonumber \\
 & & {} + H_0(\mathrm{conv}) \frac{\textrm{d} T}{\textrm{d} m} \frac{1}{T} 
  \left [ 1 - \frac{9 D}{D + \nabla - \nabla_{\mathrm{ad}}} \right .
\nonumber \\
& & \left . {} + 
  \frac{3}{2(\nabla - \nabla_{\mathrm{ad}})} \frac{\textrm{d} \nabla}
                                                  {\textrm{d} m}
  \left ( 1 + \frac{D}{D + \nabla - \nabla_{\mathrm{ad}}} \right )
\right ]
\end{eqnarray}
and
\begin{equation}
\label{eq:delta_r_deq_cc}
c_0 = \mathcal{H} - H_0(\mathrm{rad}) - H_0(\mathrm{conv}).
\end{equation}
In Eq.~\ref{eq:delta_r_deq_ac} and Eq.~\ref{eq:delta_r_deq_bc} the 
$\nabla$ is
\begin{equation}
\nabla = \frac{\textrm{d} \ln T}{\textrm{d} \ln P},
\end{equation}
$D$ is from \citet{1970SAOSR.309.....K}, and in Eq.~\ref{eq:delta_r_deq_bc} 
the $H_0(\nu)$ is the radiative flux.  Solving 
Eq.~\ref{eq:delta_r_deq_sol} for $\delta m$, using the coefficients in 
Eq.~\ref{eq:delta_r_deq_ac}, Eq.~\ref{eq:delta_r_deq_bc} and 
Eq.~\ref{eq:delta_r_deq_cc}, the corresponding temperature change based 
on conserving the flux is 
\begin{equation}
\label{eq:delta_t_flux}
\delta T_{\mathrm{flux}} = \frac{\partial T}{\partial m} \delta m.
\end{equation}

\textsc{Atlas} uses two additional temperature corrections near the 
surface, where the flux error loses sensitivity.  One correction is 
based on the flux derivative.  Because this correction applies high in 
the atmosphere where the gas is quite transparent, radiation will carry 
almost all the energy, and it is a good approximation to ignore 
convective energy transport.  The zeroth angular moment of 
the spherical radiative transfer equation (Eq.~\ref{eq:sph_rad_tran_m})
is
\begin{equation}
\label{eq:sph_moment_0}
\frac{\partial [r^2 H(\nu)]}{\partial m}
= r^2 k(\nu) [J(\nu) - S(\nu)].
\end{equation}
Replacing $J(\nu)$ by $\Lambda[S(\nu)]$, expanding the Planck function 
in $S(\nu)$ in terms of $T$, integrating over frequency and retaining 
just the diagonal terms of the $\Lambda$ operator, the resulting 
temperature correction becomes 
\begin{equation}
\label{eq:delta_t_lambda}
\delta T_{\Lambda} = 
\frac{ \frac{1}{r^2}
       \left \{ \frac{\partial(r^2 \cal{H}) } {\partial m} 
        - \frac{\partial [r^2 H(rad)] } {\partial m} \right \} }
{\int_0^{\infty} k(\nu) \frac{[\Lambda_{\mathrm{dia}}(\nu) -1] }
{[1 - \sigma(\nu) \Lambda_{\mathrm{dia}(\nu)}]}
[1 - \sigma(\nu)]
\frac{\partial B(\nu, T)}{\partial T} \textrm{d} \nu}.
\end{equation}
The term $\Lambda_{\mathrm{dia}}$ is approximated by the plane-parallel
expression given in \citet{1970SAOSR.309.....K}, assuming it has 
minimal dependence on the geometry.

A third temperature correction is used in the original \textsc{Atlas} 
code to smooth the region of overlap between the first two corrections.
This is 
\begin{equation}
\label{eq:delta_t_surf}
\delta T_{\mathrm{surf}} = \frac{T \delta H}{4 \cal{H}},
\end{equation}
and this is retained here.  The total temperature correction is 
therefore,
\begin{equation}
\label{eq:delta_t}
\delta T = \delta T_{\mathrm{flux}} + 
           \delta T_{\Lambda} + 
           \delta T_{\mathrm{surf}}.
\end{equation}

\section{Comparisons between \textsc{SAtlas\_ODF} and 
\textsc{Atlas\_ODF}}

\begin{figure}
 \resizebox{\hsize}{!}{\includegraphics{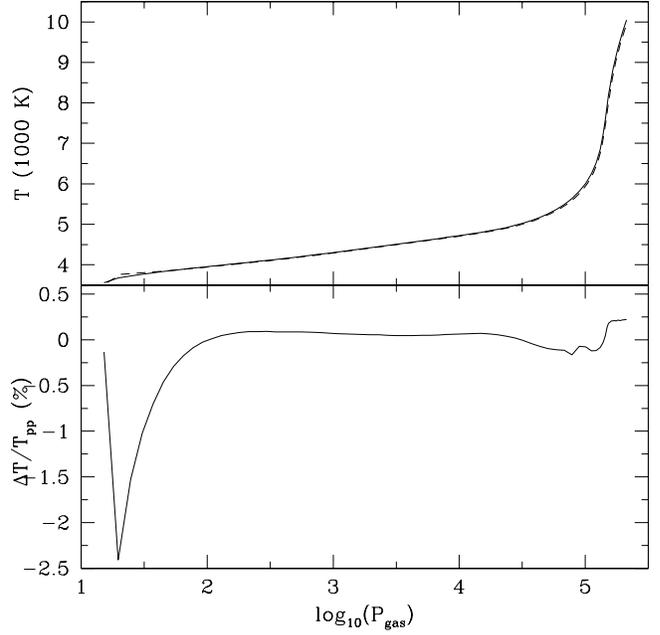}}
 \caption{\label{fig:spsol_ppsol}
 A comparison of the temperature structures for two solar atmospheric 
 models.  The top panel displays the temperatures as a function of 
 $\log_{10} (P_{\mathrm{gas}})$.  The solid line is computed using 
 spherical geometry and the dashed line is using the traditional 
 plane-parallel geometry.  The radiative transfer and the pressure 
 structures of both models were computed using the same numerical 
 techniques.  The bottom panel shows the percentage difference between 
 the temperatures of the two models, also as a function of the gas 
 pressure.}
\end{figure}

One test of the validity of the spherical code is to compute a 
spherical solar atmosphere, which should be nearly identical to the 
plane-parallel model.  For both models we used the Kurucz file 
\texttt{asunodfnew.dat} for the starting model and the file 
\texttt{bdfp00fbig2.dat} for the opacity distribution function.
To eliminate other possible sources of differences, we used the 
Bulirsch-Stoer solution to solve for the pressure structures and the 
Rybicki method for the radiative transfer in the plane-parallel model 
as well as for the spherical calculations.  The spherical model used 
the atmospheric parameters $L_{\sun} = 3.8458 \times 10^{33}$ ergs/s, 
$M_{\sun} = 1.9891 \times 10^{33}$ g, and 
$R_{\sun} = 6.95508 \times 10^{10}$ cm.  These correspond to 
$T_{\mathrm{eff}} = 5779.5$ K and $\log g = 4.43845$, which are 
slightly different from the canonical values used by Kurucz.  Therefore,
we computed the  plane-parallel model with the consistent values of 
$T_{\mathrm{eff}}$ and $\log g$.  The comparison is shown in 
Fig.~\ref{fig:spsol_ppsol}.

The $| \Delta T |$ is $\leq 0.25\%$ until 
$\log_{10} P_{\mathrm{gas}} < 2$, where the temperature of the 
spherical model begins to trend lower than the plane-parallel model.  
The dip in the temperature difference down to $- 2.4\%$ is due to the 
kink in the temperature structure of the \emph{plane-parallel} model.  
This feature was discussed earlier in connection with 
Fig.~\ref{fig:tp_ryb_josh}.  However, in this case the 
\citet{1971JQSRT..11..589R} method is used to compute the radiative 
transfer in \emph{both} models, using the \emph{same} surface boundary 
condition in both codes.  Therefore, this difference cannot be due to a 
coding difference between the two routines.  This feature might be due 
to the number of rays used in the calculation of the radiative transfer.
The Rybicki solution for the plane-parallel code uses three rays for 
each depth, whereas the same method in the spherical code uses 
$\approx 80$ rays for the layers approaching the surface.  Perhaps 
this finer griding produces a smoother temperature profile in these 
layers.  There is, however, no physical significance to the temperature 
differences near the surface because these layers are located in the 
solar chromosphere \citep{2006ApJ...639..441F}, well above the 
temperature minimum, where other physics is completely dominant.

\begin{figure}
 \resizebox{\hsize}{!}{\includegraphics{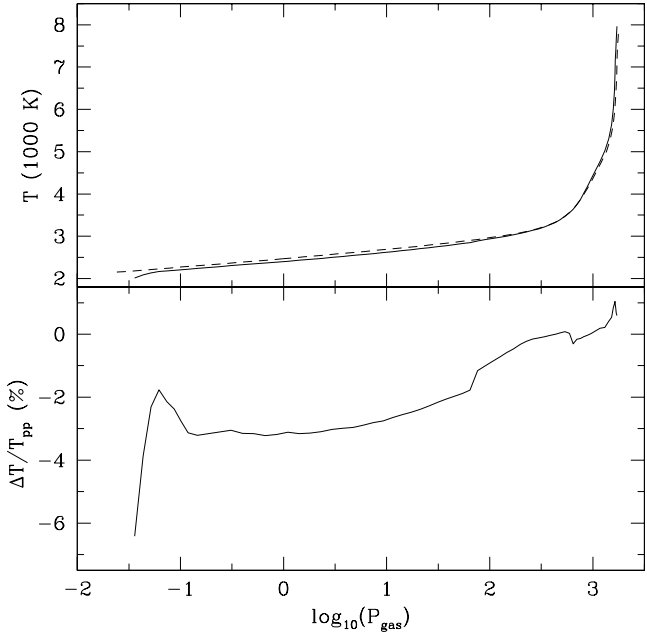}}
 \caption{\label{fig:sprg_pprg}
 A comparison of the temperature structures of two red giant atmospheric
 models.  The top panel displays the temperatures as a function of 
 $\log_{10} (P_{\mathrm{gas}})$.  The solid line is a spherical model 
 with the parameters $L = 3690 \ L_{\sun}$, $M = 1 \ M_{\sun}$ and 
 $R = 166 \ R_{\sun}$.  These values were chosen to match the 
 parameters of the plane-parallel model having 
 $T_{\mathrm{eff}} = 3500$ K and $\log g = 0.0$, computed using 
 \textsc{Atlas\_ODF}, shown by the dashed line. 
 The bottom panel shows the percentage difference between the 
 temperatures of the two models, also as a function of the gas 
 pressure.}
\end{figure}

A test where larger differences are expected is for the coolest model 
($T_{\mathrm{eff}} = 3500$ K, $\log g = 0.0$) in the grid 
\texttt{/grids/gridp00odfnew/ap00k2odfnew.dat} (computed by Castelli) 
on the Kurucz web site.  Because $T_{\mathrm{eff}}$ and $\log g$ really 
represent the three parameters $L$, $M$ and $R$, there is a degeneracy 
that must be broken.  To do this, we have assumed the star has 
$M = 1 \ M_{\sun}$, which leads to $L_{\ast} = 3690 \ L_{\sun}$ and 
$R = 166 \ R_{\sun}$.  The comparison of the atmospheric structures is 
shown in Fig.~\ref{fig:sprg_pprg}.  While the plane-parallel model was 
taken from the grid on Kurucz's web site, that model served only as the 
starting point for computing the model structure using our 
\textsc{Atlas\_ODF} code to ensure that it reflects the same numerical 
routines.  In particular, we used the Rybicki routine for the radiative 
transfer of both the plane-parallel and the spherical models so that 
the resulting differences must come from the atmosphere's geometry, and 
not the method of solution.

The spherical model is cooler than the plane-parallel model throughout 
most of the atmosphere, and it becomes progressively cooler with 
increasing height.  The distance from $r(1)$ to 
$r(\tau_{\mathrm{R}} = 2/3) = R_{\ast}$ is $2.12 \times 10^7$ km, 
giving an atmospheric extension, defined in section 3.1, of 0.18.  In 
the plane-parallel model the corresponding distance 
$d(1)$ - $d(\tau_{\mathrm{R}} = 2/3) = 1.85 \times 10^7$ km.  Deep in 
the atmosphere the spherical model becomes systematically hotter than 
the planer-parallel model as the core makes a greater contribution.

\begin{figure}
 \resizebox{\hsize}{!}{\includegraphics{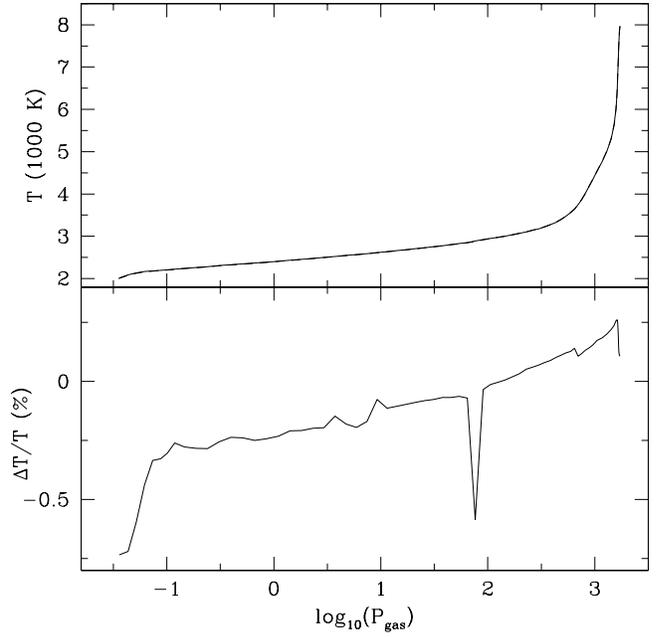}}
 \caption{\label{fig:sprg_sp2rg}
 A comparison of the structures of two spherical red giant atmospheric 
 models, both equivalent to $T_{\mathrm{eff}} = 3500$ K and 
 $\log g = 0.0$.  The top panel displays the temperatures as a function 
 of $\log_{10} (P_{\mathrm{gas}})$.  The solid line is a spherical 
 model having $L = 3690 \ L_{\sun}$, $M = 1 \ M_{\sun}$ and 
 $R = 166 \ R_{\sun}$.  The dashed line, which is nearly coincident 
 with the solid line, represents the model having the parameters 
 $L = 2952 \ L_{\sun}$, $M = 0.8 \ M_{\sun}$ and $R = 148 \ R_{\sun}$.
 The bottom panel shows the percentage difference between the 
 temperatures of the two models, also as a function of the gas pressure.
 The sense of the differences is the less massive and luminous model 
 minus the more massive and luminous model.
 }
\end{figure}

Because of the degeneracy of the atmospheric parameters, we tried 
another combination of luminosity, mass and radius that also match 
$T_{\mathrm{eff}} = 3500$ K and $\log g = 0.0$, namely, 
$L_{\ast} = 2952 \ L_{\sun}, M = 0.8 \ M_{\sun}$ and 
$R = 148 \ R_{\sun}$.  The comparison of the two spherical models, both 
computed with \textsc{SAtlas\_ODF}, is shown in 
Fig.~\ref{fig:sprg_sp2rg}.  The structures are so similar that they 
seem identical in the top panel.  In the bottom panel, where the 
differences in the structures are plotted, it is easier to see that the 
less massive and luminous star has a slightly steeper temperature 
profile, as is expected.

\section{Comparison with other programs}

\begin{figure}
 \resizebox{\hsize}{!}{\includegraphics{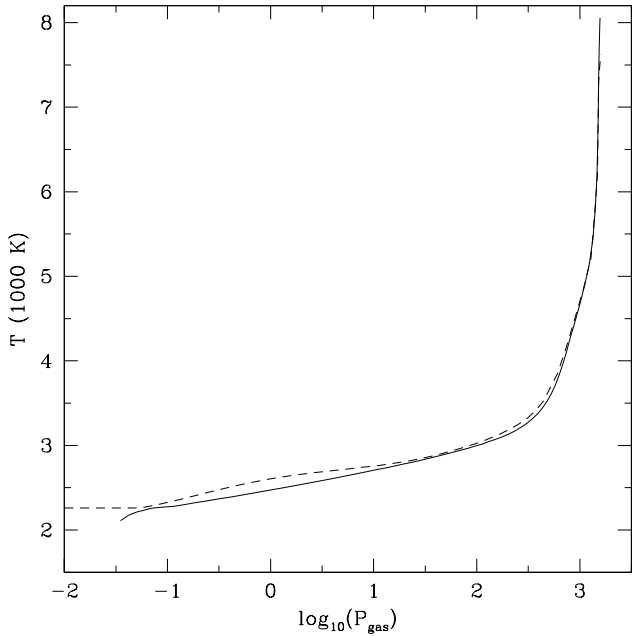}}
 \caption{\label{fig:as_ng}
          A comparison of the temperature structures of an 
          \textsc{SAtlas\_ODF} atmosphere (solid line) with the 
          structure of a model from the NG-giant grid computed with the 
          \textsc{Phoenix} code (dashed line).  The \textsc{SAtlas}
          model has the atmospheric parameters $L = 10324 \ L_{\sun}$, 
          $M = 2.5 \ M_{\sun}$ and $R = 262 \ R_{\sun}$, corresponding
          to the NextGen parameters $T_{\mathrm{eff}} = 3600$ K, 
          $\log g = 0.0$ and $M = 2.5 \ M_{\sun}$.}
\end{figure}

The \textsc{Phoenix} program \citep{1999ApJ...512..377H} can also 
compute LTE, line-blanketed, spherically extended models, and a 
comparisons with those models is appropriate.  The \textsc{Phoenix} web 
site 
(\textit{http://www.hs.uni-hamburg.de/EN/For/ThA/phoenix/index.html})
contains the NG-giant grids, in which the model that is closest to the 
examples used above is the one with $T_{\mathrm{eff}} = 3600$ K, 
$\log g = 0.0$ and $M = 2.5 \ M_{\sun}$.  To compare with this model, 
we have computed a spherical model with 
$L = 10324 \ L_{\sun}, \ M = 2.5 \ M_{\sun}$ and 
$R = 262 \ R_{\sun}$, again starting from the same plane-parallel model 
with $T_{\mathrm{eff}} = 3500$ K and $\log g = 0.0$ that we used 
earlier.  The comparison is shown in Fig.~\ref{fig:as_ng}.  Now the 
differences are somewhat larger than in the previous comparisons, which 
is to be expected because the detailed calculations are nearly totally 
independent.  Overall, however, the comparison is very similar, showing 
that the two models have essentially the same structures, although we 
note that the NextGen model has a temperature bulge compared with our 
model in the pressure range $-1 < \log_{10} P_{\mathrm{gas}} < + 1.5$.
We have not observed this kind of feature in the comparisons we have 
made with the Kurucz models or between our spherical and plane-parallel
models.

A note about the relative run times of the two codes: the time per 
iteration running \textsc{SAtlas\_ODF} on our single-processor desktop 
workstation is just 5\% of the time per iteration given in the header 
files of the NG-giant model.

\begin{figure}
 \resizebox{\hsize}{!}{\includegraphics{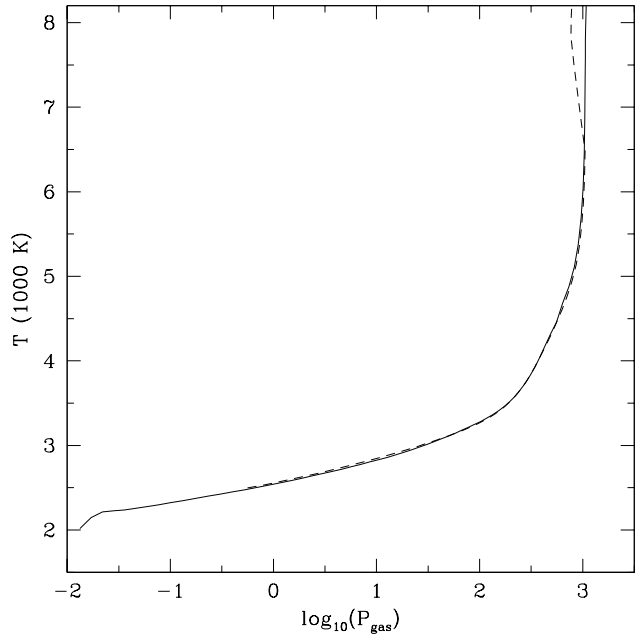}}
 \caption{\label{fig:as_m}
          A comparison of the temperature structures of an 
          \textsc{SAtlas\_ODF} atmosphere (solid line) with a 
          spherical \textsc{MARCS} model (dashed line).
          The \textsc{SAtlas} model has the atmospheric parameters 
          $L = 6390 \ L_{\sun}$, $M = 1.0 \ M_{\sun}$ and 
          $R = 166 \ R_{\sun}$, corresponding to the \textsc{MARCS}
          parameters $T_{\mathrm{eff}} = 4000$ K, 
          $\log g = 0.0$ and $M = 1.0 \ M_{\sun}$.}
\end{figure}

The \textsc{MARCS} program \citep{2008A&A...486..951G} is another well 
established code that has the ability to compute LTE, line-blanketed, 
spherical model atmospheres.  From the MARCS web site 
(\textit{http://marcs.astro.uu.se/}) the model with parameters 
$T_{\mathrm{eff}} = 4000$ K, $\log g = 0.0$ and $M = 1 \ M_{\sun}$ is
closest the the examples we have been using.  This model also has 
solar abundances and a microturbulent velocity of 2 km/s.  The header 
lines in the model gives the spherical parameters 
$L = 6390 \ L_{\sun}$ and 
$R = 1.1550 \times 10^{13} \ \textrm{cm} = 166 \ R_{\sun}$. 
MARCS defines the radius at $\tau_{\mathrm{R}} = 1.0$, not at 
$\tau_{\mathrm{R}} = 2/3$ that we use, but this is a small difference.
Therefore, we have started from the model with 
$T_{\mathrm{eff}} = 4000$ K, $\log g = 0.0$ and 
microturbulence = 2 km/s in the same grid 
(\texttt{/grids/gridp00odfnew/ap00k2odfnew.dat}) used earlier, and we 
have computed a spherical model with the luminosity, mass and radius of 
the MARCS model.  The comparison is shown in Fig~\ref{fig:as_m}. 
The models agree very well in the range 
$\log_{10} P_{\mathrm{gas}} < + 1.5$, where the NextGen model
displayed a temperature bulge.  But, the spherical MARCS model appears 
to have a pressure inversion at $T > 6500$ K, something that is not 
present in the comparison with the \textsc{Phoenix} model.  Overall,
however, the structures are in substantial agreement.

\section{Conclusions}

We have modified the robust, open-source, plane-parallel model 
atmosphere program \textsc{Atlas} to treat spherically extended 
geometry.  The resulting spherical code, \textsc{SAtlas}, which is 
available in both opacity distribution function and opacity sampling 
versions, was used to compute several test models.  At high surface 
gravity the spherical model structure is essentially identical to the 
plane-parallel model structure.  At low surface gravity, the 
\textsc{SAtlas} models agree very well with the spherical model 
structures computed by \textsc{Phoenix} and by \textsc{MARCS}.   The 
\textsc{SAtlas} program, which runs easily on a desktop workstation, 
offers a viable alternative for modeling the atmospheres of low surface 
gravity stars.

As an example of the utility of \textsc{SAtlas}, we have used it to 
compute more than 2500 models to create model cubes with fine parameter 
spacing covering the specific $L_{\ast}$, $M_{\ast}$ and $R_{\ast}$ 
values needed for an analysis of the optical interferometry of just 
three stars (Neilson \& Lester submitted).  We are in the process of 
computing more models for our own application.  These codes and models 
are available at 
\texttt{http://www.astro.utoronto.ca/$\sim$lester/Programs/}.

\textit{Acknowledgments} This work is built on the development of the 
original \textsc{Atlas} programs by Robert Kurucz, and his 
generosity in making his source codes, line lists and opacity 
distribution functions freely available.  Our modifications have 
benefited greatly from the many times that he has answered our 
questions and from the test models he has provided.  We also  
gratefully acknowledge the comments and sample models provided by 
Fiorella Castelli that have aided our efforts.  We also thank the 
anonymous referee for asking numerous questions that prompted us to 
provide more thorough explanations of our results.

This work has been supported by a research grant from the Natural 
Sciences and Engineering Research Council of Canada.  HRN has received 
financial support from the Walter John Helm OGSST and the Walter C. 
Sumner Memorial Fellowship.

\bibliographystyle{aa} 
\bibliography{0578}

\pagebreak
 
\end{document}